\documentclass[12pt]{iopart}
\pdfoutput=1 
\usepackage[utf8]{inputenc}
\expandafter\let\csname equation*\endcsname\relax
\expandafter\let\csname endequation*\endcsname\relax
\usepackage{amsmath}
\usepackage{amssymb}
\usepackage{graphicx}
\usepackage{dcolumn}
\usepackage{bm}
\usepackage{appendix}
\usepackage{xcolor}
\usepackage{hyperref}
\hypersetup{
    colorlinks=true,
    linkcolor=purple,
    citecolor=teal
    }
\usepackage{ulem}
\graphicspath{{figures/}}
\newcommand{\bx}{\mathbf{x}}

\begin{document}
\title{Non-reciprocal interactions spatially propagate fluctuations in a 2D Ising model}

\author{Daniel S. Seara}
\address{James Franck Institute, University of Chicago, Chicago IL, USA}
\author{Akash Piya}
\address{Thomas Jefferson High School for Science and Technology, Alexandria VA, USA}

\author{A. Pasha Tabatabai}
    \ead{To whom correspondence should be addressed: tabatabai@seattleu.edu}
\address{Department of Physics, Seattle University, Seattle WA, USA}

\begin{abstract}
    Motivated by the anisotropic interactions between fish, we implement spatially anisotropic and therefore non-reciprocal interactions in the 2D Ising model. First, we show that the model with non-reciprocal interactions alters the system critical temperature away from that of the traditional 2D Ising model. Further, local perturbations to the magnetization in this out-of-equilibrium system manifest themselves as traveling waves of spin states along the lattice, also seen in a mean-field model of our system. The speed and directionality of these traveling waves are controllable by the orientation and magnitude of the non-reciprocal interaction kernel as well as the proximity of the system to the critical temperature.
    
\end{abstract}

\maketitle

\section{Introduction}
When two objects come into contact with each other, the macroscopic forces that each generate on the other are described by Newton's Third Law and are equal in magnitude. While the reciprocity of this type of interaction is common, there are many instances where interactions between objects are non-reciprocal and lead to interesting behaviors. For example, metamaterials that exhibit broken symmetries in the bonds between constituents yield asymmetric responses to mechanical~\cite{coulais_static_2017, nassar_non-reciprocal_2017, brandenbourger_non-reciprocal_2019} and optical waves~\cite{lin_unidirectional_2011, ramos_optical_2020} as well as fluid/solid behavior~\cite{tabatabai_detailed_2021, scheibner_odd_2020}. Non-reciprocal interactions are also used as design principles for sensor optimization~\cite{ngampruetikorn_energy_2020}.

Hallmark examples of non-reciprocal interactions occur within the collective behavior of animal groups such as locusts, birds, and fish. In these systems, visual information differences between animals lead to this non-reciprocal interaction. While incorporating non-reciprocal interactions are not required to capture flocking behavior~\cite{vicsek_novel_1995}, there is a recent push towards understanding the effects of non-reciprocal interactions on phase behavior~\cite{durve_first-order_2016, barberis_large-scale_2016, durve_active_2018} and the non-trivial motion of flocking objects~\cite{lecheval_social_2018}.

In particular, we are interested in how fluctuations in the local polarization within a flock are propagated through space as a consequence of these non-reciprocal interactions. To this end, we simplify the problem and study non-reciprocal interactions within a 2D Ising model that are motivated by the anisotropic field of view within animals such as fish, consistent with recent efforts to understand the effects of
non-reciprocal interactions within the continuum Vicsek model ~\cite{durve_first-order_2016, durve_active_2018}. This model is not equivalent to the active Ising model~\cite{solon_flocking_2015,yu_energy_2022} since objects are not free to move on the lattice and lack self-propulsion. As a consequence, our model decouples the non-equilibrium effects of introducing a non-reciprocal interaction and active energy consumption. We note that most non-reciprocal interactions within studies of flocking focus on so-called `vision-cones' where object orientation influences interactions. Our interaction is a simplified version of a vision-cone which does not change orientation.

Our results are presented in three parts. First, the model is introduced and the influence of the non-reciprocal interaction on the phase behavior of the system is described in detail. Comparisons are made to the equilibrium 2D Ising model. Then, we characterize the spatial propagation of fluctuations that originate from the presence of this non-reciprocal interaction. We introduce a mean-field model which predicts the propagation of spin fluctuations under these parity-breaking interactions. We show that this propagation is robust to changes in the algorithm used to generate the system dynamics. Finally, we find that the propagation velocity is maximized near the critical temperature.

\section{Methods}
\begin{figure}
    \centering
    \includegraphics[width=.8\columnwidth]{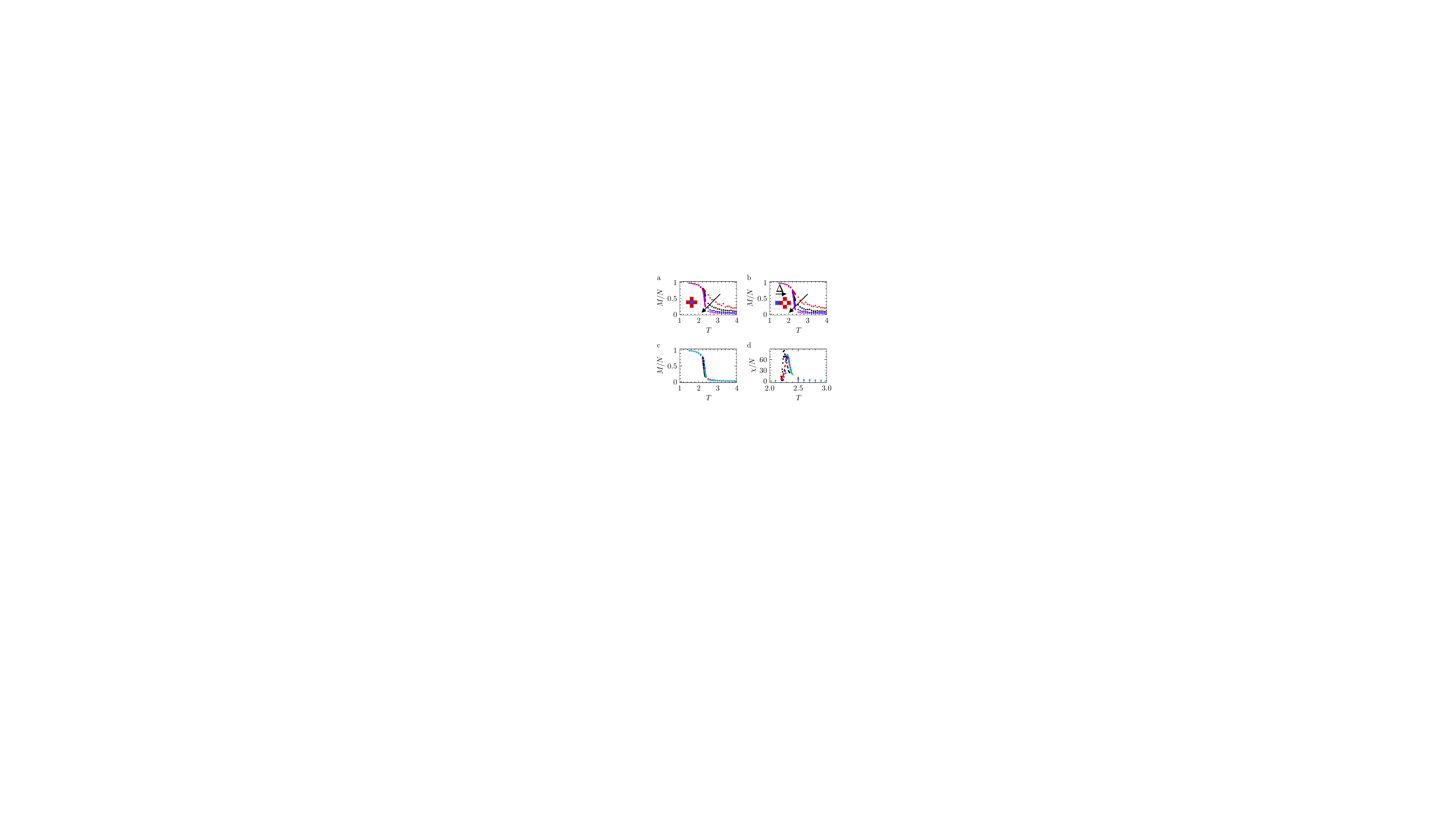}
    \caption{\textbf{Size and Offset Effects on System Magnetization.} (a) Magnetization per spin $M/N$ as a function of temperature $T$ for lattices with $L=8$ (red), $L=16$ (black), $L=32$ (blue), and $L=64$ (magenta). (a-inset) Representation of a standard Ising Model interaction kernel. Neighbors of the blue cell are defined as the four red lattice sites. (b) Magnetization per spin as a function of temperature for offset $\Delta =2$ with symbol colors equivalent to (a). (b-inset) Representation of an non-reciprocal interaction kernel with an offset  $\Delta = 2$ lattice sites. Arrows point in the direction of increasing system size. (c) Magnetization per spin $M/N$ as a function of temperature $T$ for lattices of size $L=64$ at different values of offset $\Delta=0$ (red), $\Delta = 2$ (black), $\Delta = 4$ (blue), $\Delta = 8$ (magenta), and $\Delta = 10$ (cyan). (d) Magnetic susceptibility per spin $\chi/N$ for data in (c). } 
    \label{fig:fig1}
\end{figure}

We consider a 2D Ising model on a square lattice of dimension $L\times L$ with $N=L^2$ spins where the system at a time $t$ is in a configuration denoted by the vector $\mathbf{s}(t)$ with elements $s_i =  \pm 1$, where $i \in \mathbb{Z}^2$ indexes the lattice site~\cite{onsager_crystal_1944}. In order to generate spin dynamics on the lattice, we use
the energy of an individual spin $s_i$, calculated as \cite{Sanchez2002, Lima2006, Lipowski2015}
\begin{equation}
    \label{eq:spinEnergy}
    E_i(\mathbf{s}) = -s_i \sum_j J_{ij} s_j
\end{equation}

where $J_{ij}$ is an element of an interaction matrix which couples spins $i$ and $j$.
Note that this is related, but distinct, from the traditional Ising model Hamiltonian, given by $E = (\sum_i E_i) / 2$ \cite{Kardar2007}. Importantly, while $E_i(\mathbf{s})$ is not an energy functional, we nevertheless have $\Delta E = \Delta E_i$ when $J_{ij} = J_{ji}$ (see \ref{app:MC_methods} for details). When this symmetry condition on the interaction matrix is satisfied, we call the system reciprocal. Otherwise, if $J_{ij} \neq J_{ji}$, we call the system non-reciprocal \cite{fruchart_non-reciprocal_2021}.

To begin, we study an equilibrium reciprocal Ising model where spins $i$ and $j$ interact with strength $J_{ij} =J$ if they are nearest neighbors, otherwise $J_{ij}=0$ (Fig.~\ref{fig:fig1}a-inset). We can write the interaction matrix as
\begin{equation}
    J_{ij} = J 
    (
    \delta_{i - \hat e_x, j} + \delta_{i + \hat e_x, j} +
    \delta_{i - \hat e_y, j} + \delta_{i + \hat e_y, j}
    ).
\end{equation}
In the above, $\hat{e}_\alpha$ denotes the unit vector in the $\alpha$ direction, with $\alpha \in \lbrace x, y \rbrace$. Unless otherwise stated, we set $J=1$.

Lattices are placed in contact with a heat bath at a temperature $T$, with units of the Boltzmann constant $k_B$ (Supplemental Movie 1). Assuming periodic boundary conditions, the lattices are evolved towards their thermodynamic equilibrium configurations using a Metropolis Monte Carlo method (\ref{app:MC_methods}). Briefly, individual spins in the lattice are chosen at random and the energy cost/gain of flipping the spin as given by the variation of Eq.~\ref{eq:spinEnergy} determines the probability of flipping the spin 
as shown by
\begin{equation}
    P(\mathbf{s} \to F_i \mathbf{s}) = \mathrm{min} \left[ 1, e^{\beta \Delta E_i} \right]
\end{equation}
where $F_i$ is an operator that takes $s_i \mapsto -s_i$ and $\Delta E_i$ is given by Eq.~\ref{eq:spinEnergyChange} in \ref{app:MC_methods}~\cite{giordano_computational_2006}.

We define a `sweep' as a proxy for time; in a single sweep of a lattice with $N$ lattice sites, $N$ sites are randomly selected sequentially and have the possibility of flipping their spin. Systems are brought to a steady-state configuration and ensemble measurements of the system magnetization are made of independent configurations (\ref{app:MC_Measure}).

For a given lattice, we calculate the system 
magnetization from the sum of all spins $s_i$, $M=\sum_i s_i$, which exhibits the expected temperature and system-size dependence previously described (Fig.~\ref{fig:fig1}a)~\cite{giordano_computational_2006}.
Since the $\mathbf{s} \mapsto -\mathbf{s}$ symmetry is not broken by the existence of an external field, we quote the absolute value of the magnetization.

Next, we amend the traditional Ising model interaction matrix and introduce a non-reciprocal interaction ($J_{ij} \neq J_{ji}$) in Eq.~\ref{eq:spinEnergy}. Inspired by the non-local information processing of active systems, such as fish within a school or starlings within a flock~\cite{couzin_collective_2002, katz_inferring_2011, ballerini_interaction_2008, durve_first-order_2016}, we define an offset vector $\boldsymbol{\Delta} \in \mathbb{Z}^2$ that represents a spatial translation of the interaction kernel (Fig.~\ref{fig:fig1}b-inset),
\begin{equation}
    \label{eq:offsetVector}
    \boldsymbol{\Delta} = \Delta_x \hat e_x + \Delta_y \hat e_y,
\end{equation}
where $\Delta_\alpha \in \mathbb{Z}$ are integers. The interaction matrix's elements are now given by
\begin{equation}
\begin{split}\label{eq:bigJ}
    J_{ij} = J(& \delta_{i + \boldsymbol{\Delta} - \hat e_x, j} + \delta_{i + \boldsymbol{\Delta} + \hat e_x, j} + \delta_{i + \boldsymbol{\Delta} - \hat e_y, j} + \delta_{i + \boldsymbol{\Delta} + \hat e_y, j}
    )
\end{split}
\end{equation}
We choose to keep $\boldsymbol{\Delta}$ constant and uniform throughout space, specifically $\boldsymbol{\Delta} \equiv \Delta \hat e_x$, and $\Delta$ is not dependent on the sign of the lattice spin. These unidirectional interactions lead to qualitatively similar temperature and size-dependence of the system magnetization (Fig.~\ref{fig:fig1}b). Note that the traditional Ising model is represented by $\Delta =0$, and we keep a constant number of interactions between spins (4) to keep the connectivity of the lattice constant for all values of $\Delta$.

We continue to use the changes in $E_i$ in Eq.~\ref{eq:spinEnergy} when calculating the transition probabilities in the Monte Carlo simulations to generate the system dynamics, although we again note that it is no longer the case that $\Delta E = \Delta E_i$.
The incorporation of asymmetrical interactions in Eq.~\ref{eq:spinEnergy} to generate lattice dynamics has been used in previous studies of Ising models on directed graphs~\cite{Sanchez2002, Lima2006, Lipowski2015}. Further, directed Ising models with asymmetric interactions have been described analytically in 1D and in 2D within the zero temperature and paramagnetic limits~\cite{Godreche2011, Godreche2014, Godreche2015}.

The qualitative temperature dependence of the system magnetization is similar for all values of $\Delta$, however subtle differences are observed for different values of $\Delta$ (Fig.~\ref{fig:fig1}c). To explore these subtle differences, we calculate the variance of the magnetization $\Delta M$ over time, where $(\Delta M)^2 = \left<M^2\right> - \left<M\right>^2$, to calculate the  magnetic susceptibility $\chi = (\Delta M)^2 /T$~\cite{giordano_computational_2006} \footnote{We note that this susceptibility is no longer guaranteed to be derivable from the individual energy used to generate the system dynamics, i.e. $\chi \neq \lim_{h \to 0}\left( \frac{\partial^2 E_i}{\partial h^2} \right)\Big\vert_{T}$ \cite{Kardar2007}.}. The change in location of the peak of $\chi$ suggests that $\Delta$ may influence the critical temperature of the system away from the expected critical temperature for the 2D Ising model on a square lattice, $T_c = 2 / \ln(1 + \sqrt{2}) \approx 2.27$ (Fig.~\ref{fig:fig1}d).

\section{Results}
\begin{figure}
    \centering
    \includegraphics[width=.8\columnwidth]{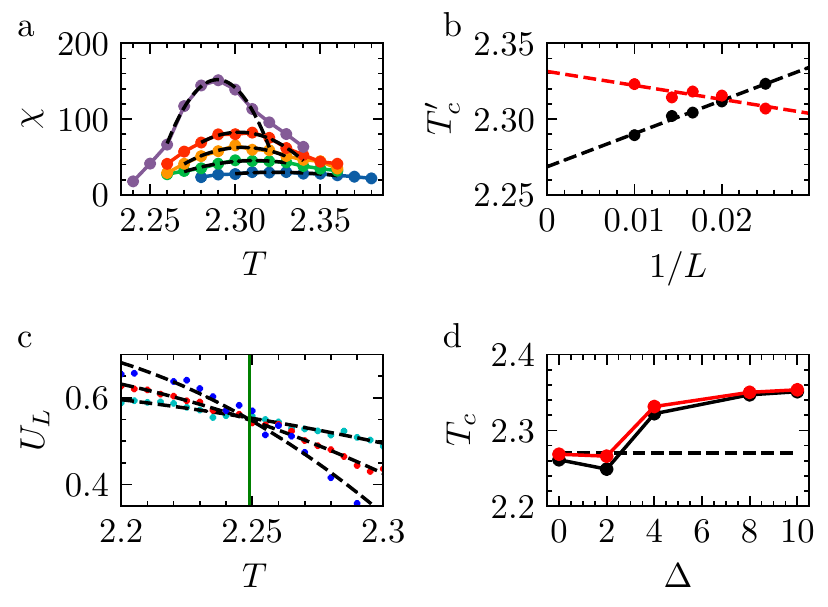}
    \caption{\textbf{Critical Temperature Depends on $\Delta$.} (a) Susceptibility $\chi$ as a function of temperature $T$ for $\Delta =0$ on lattices of length $L=40,50,60,70,100$ shown in blue, green, orange, red, and purple, respectively. Dashed black lines are quadratic fits near the maxima. (b) Finite size critical temperature $T_c'$ as a function of inverse lattice size $1/L$ for $\Delta = 0$ (black) corresponding to data in (a) and $\Delta = 4$ (red). Dashed lines are linear fits. (c) Binder cumulant $U_L$ as a function of temperature $T$ for $\Delta =2$ and system size $L=16,32,64$ in cyan, red, and blue respectively. Dashed lines are second-order polynomial fits, and the green vertical line indicates the intersection of the fitted lines. (d) Infinite system size critical temperature $T_c$ as a function of offset $\Delta$ calculated by extrapolation of susceptibility peak (red) and Binder cumulant intersection (black). Red data error bars representing the standard error are plotted but are smaller than the symbols. Dashed line at the equilibrium $T_c = 2.7$ for reference.}
    \label{fig:Tc_measure}
\end{figure}

We use two methods to determine the infinite system size critical temperature $T_c$ to determine if non-reciprocal interactions affect $T_c$. First, we use the temperature which gives a maximum in $\chi$ to define the finite-size critical temperature $T_c'$ for a lattice of length $L$ (Fig.~\ref{fig:Tc_measure}a, Supplemental Fig. 1)~\cite{ferdinand_bounded_1969}. We determine the maximum of $\chi(T)$ by fitting $\chi(T)$ to a second-order polynomial. We then use $T_c'$ for lattices of different sizes to determine the infinite system size critical temperature $T_c$ which is found from the y-intercept in Fig.~\ref{fig:Tc_measure}b. We find that $T_c$ changes over the range of $\Delta$ studied despite a constant connectivity of each lattice. Interestingly, the slopes of $T_c'(1/L)$ for $\Delta \neq 0$ are negative, whereas the slope $T_c'(1/L)$ for $\Delta=0$ is positive.

As an orthogonal approach to confirm these values of $T_c$, we calculate the Binder cumulant $U_L=1-\left< M^4\right>/3\left<M^2\right>^2$ for systems of size $L$ using higher order moments of the system magnetization $M$~\cite{binder_monte_1988}. The averages $\left<...\right>$ are taken over an ensemble of independent measurements for a given system size and temperature. Finite-sized scaling yields an intersection of the Binder cumulant $U_L(T)$ for systems of different sizes at the infinite system size critical temperature $T_c$ (Fig.~\ref{fig:Tc_measure}c, ~\ref{app:binder}, Supplemental Fig. 2). We find $T_c$ from this intersection by fitting $U_L(T)$ to a second-order polynomial and minimizing the difference between $U_L(T)$ for three different lattice sizes (Fig.~\ref{fig:Tc_measure}c). Again, we observe that $\Delta$ affects $T_c$ (Fig.~\ref{fig:Tc_measure}d). We note here that measurements of $T_c$ from the extrapolation of finite-size $T_c'$ via the susceptibility measurements as well as the Binder cumulant crossing methods are done for unique sets of simulated data, strengthening our conclusion that the quantitative trends of $T_c(\Delta)$ in Fig.~\ref{fig:Tc_measure}d are robust.

Using both methods of calculating $T_c$, we observe that this non-reciprocal interaction alters the system critical temperature.
Importantly, the measured $\Delta = 0$ values of $T_c=2.269\pm 0.002$ and $T_c = 2.261$ using the susceptibility peaks and Binder cumulant intersection methods, respectively, are consistent with the standard 2D Ising model~\cite{fisher_renormalization_1998, stanley_introduction_1987}. Next, we investigate the dynamical consequences of this new model.



\begin{figure}
    \centering
    \includegraphics[width=.8\columnwidth]{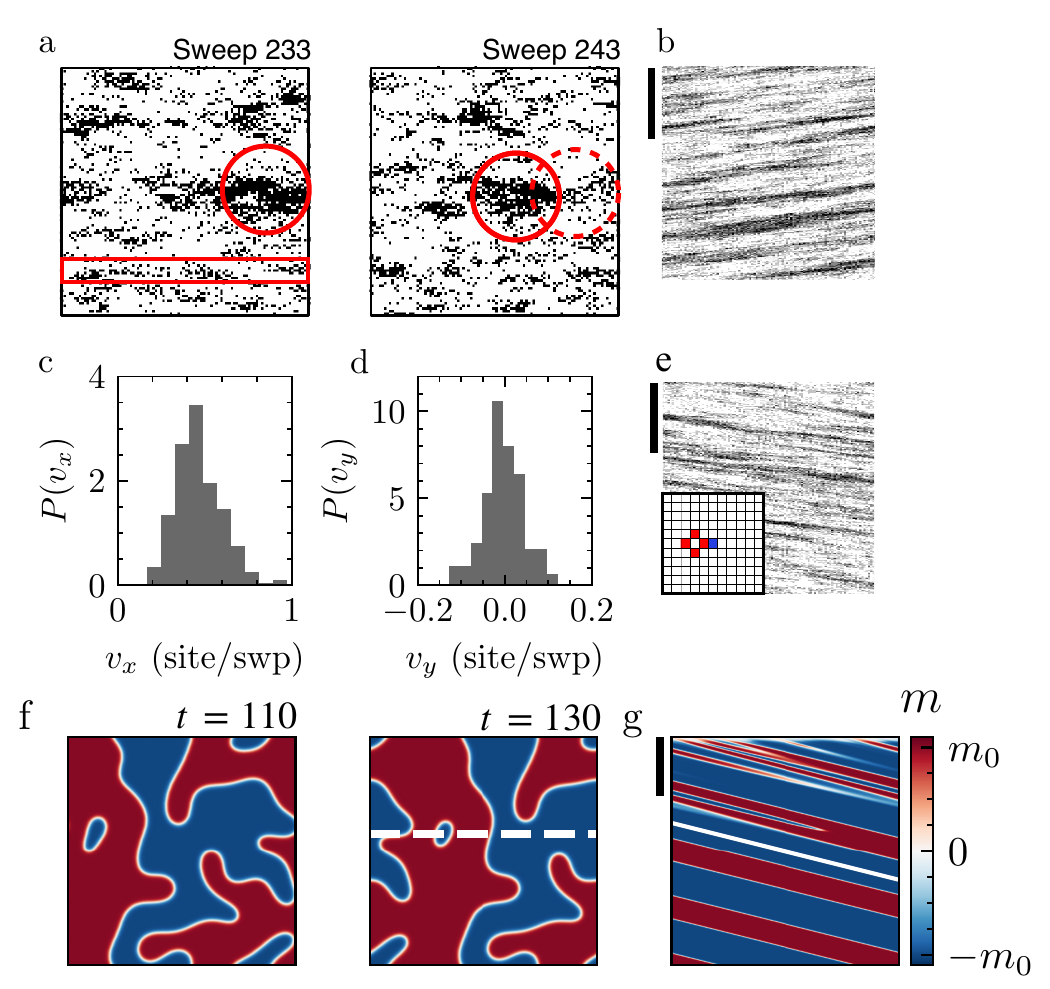}
    \caption{\textbf{Offsets Generate Flow.} (a) Temporal evolution of $L=100$ lattice at T = 2.25 and $\Delta = 2$. Images in (a) are 10 sweeps apart. The red circles highlight domain motion. Red rectangle represents region of linescan used in kymograph. (b) Kymograph of lattice in (a). Probability densities of the (c) x-components of the velocity vectors $v_x$, where $\hat{x}$ is defined to the left, and (d) y-components of the velocity $v_y$ calculated from PIV on the series represented in (a). (e) Kymograph of images created with $\Delta = -2$ (e-inset). Scale bars in (b,e) are equal to 100 sweeps. Widths of (b,e) are $L=100$. (f) Snapshots of local magnetization $m$ from mean-field theory simulations using Eq.~\ref{eq:pitchfork+advection} at different times. (g) Kymograph of the system taken at the white dashed line in (f). Time is increasing downwards and the black bar indicates $t = 50$ in simulation units. The white line shows the predicted wave velocity, Eq.~\ref{eq:meanfieldspeed}. Simulations in (f)-(g) share the colorbar shown in (g) and are run with $ J = 1$, $T = 2$, $\Delta = -1$, and $L = 100$.} 
    \label{fig:flow_example}
\end{figure}

Strikingly, we observe that systems with interactions $\Delta>0$ exhibit spin domains that appear to translate (Supplemental Movie 2). Since individual spins are constrained to their lattice site, this translation of spin domains is the propagation of local fluctuations within the magnetization which we treat as an apparent flow. These waves are non-periodic in space.

We display the dynamics of this flow in a system with an offset of $\Delta = 2$ lattice sites (Fig.~\ref{fig:flow_example}a, Fig.~\ref{fig:fig1}b-inset). We render lattices as binary images where spins +1/-1 correspond to white/black pixels respectively (\ref{app:image_analysis}). Reversing this color association does not qualitatively affect our results (Supplemental Fig. 3). After evolving the system to steady state, we observe domains that translate across the lattice. Therefore, these apparently mobile domains are reminiscent of a traveling wave. 

To demonstrate the persistence of these traveling spin waves, we build a kymograph by taking a linescan across the lattice in subsequent images (Fig.~\ref{fig:flow_example}b, \ref{app:image_analysis}); the diagonal `stripes' in Fig.~\ref{fig:flow_example}b are reminiscent of an object moving with a constant speed in a space-time diagram. To quantify the speeds of these traveling waves, we use particle image velocimetry (PIV) on time-series images (\ref{app:image_analysis}). Consistent with the kymographs, PIV reveals a non-zero wave speed $v_x$ in the direction anti-parallel to $\Delta$. Velocity distributions perpendicular to the offset $v_y$ are centered around zero (Fig.~\ref{fig:flow_example}c-d). 

We further demonstrate that the orientation of the interaction kernel dictates the orientation of the traveling wave. We set $\Delta = -2$ and confirm that the traveling wave reverses direction (Fig.~\ref{fig:flow_example}e, Supplemental Movie 3).

In order to better understand the origin of these waves, we explicitly calculate mean-field dynamics for our modified Ising model, whose form can be understood on phenomenological grounds. The equilibrium Ising model follows a pitchfork bifurcation in addition to diffusion \cite{Glauber1963, Suzuki1968, Muller1973, Cross1993, Strogatz2018}:
\begin{equation}
    \label{eq:pitchfork}
    \partial_t m(\bx, t) = a m - b m^3 + D \nabla^2 m.
\end{equation}
where $m(\bx, t)$ is a scalar field in $\bx \in \mathbb{R}^2$, $D$ is a diffusion constant, $\nabla$ is the gradient operator, $\nabla^2 = \nabla \cdot \nabla$, and $a$ and $b$ are phenomenological constants. In order to maintain stability of the homogeneous solutions, we require $b > 0$. When $a < 0$, the system has a disordered, homogeneous solution $m = 0$. When $a > 0$, the system becomes ordered, with $m = \pm m_0$, where $m_0 = \sqrt{a / b}$.

Any modification to these dynamics should reflect the explicitly broken parity symmetry introduced by the spatially uniform offset interaction kernel. To lowest order in gradients, this is accomplished by
\begin{equation}
    \label{eq:pitchfork+advection}
    \partial_t m(\bx, t) + (\mathbf v \cdot \nabla) m = a m - b m^3 + D \nabla^2 m.
\end{equation}
The new term introduced on the left-hand side, $(\mathbf v \cdot \nabla) m$, quantifies advection of $m$ in a velocity field $\mathbf v(\bx, t)$, and is precisely what is calculated in the mean-field dynamics of our modified Ising model in the $\Delta \to 0$ limit. As the interaction kernel is uniform across time and space, $\mathbf v (\bx, t) = v_0 \hat{\mathbf{e}}_{\boldsymbol{\Delta}}$, where $\hat{\mathbf{e}}_{\boldsymbol{\Delta}}$ is a unit vector pointing in the direction of the offset vector $\boldsymbol{\Delta}$ and $v_0$ is the advection speed. From the microscopic derivation, we find
\begin{equation}\label{eq:meanfieldspeed}
v_0 = -4 J \Delta/T.
\end{equation}
The relation $v_0 \propto -\Delta$ is in agreement with Monte Carlo simulations (Fig.~\ref{fig:flow_example}b,e). For comparison, snapshots of simulations of Eq.~\ref{eq:pitchfork+advection} in Fig.~\ref{fig:flow_example}f-g are qualitatively similar to the Monte Carlo simulations (\ref{app:dedalus}). Derivations and all parameters are detailed in Supplemental Note A2-3 (Supplemental Movie 4).

Our mean-field theory, Eq.~\ref{eq:pitchfork+advection}, predicts that the system becomes uniformly magnetized when $a < 0$, coarsening to a uniform state with a time-scale $L(t) \sim t^{1/2}$, as expected for non-conserved order parameters \cite{Bray1994}. In the uniform steady state, a traveling state is not defined. Nevertheless, we observe steady-state traveling waves due to the presence of domain-wall solutions to Eq.~\ref{eq:pitchfork+advection}, which propagate with velocity $v_0$. These soliton-like solutions have the form
\begin{equation}
    \label{eq:soliton}
    m(\mathbf{x}, t) = m_0 \tanh(\gamma (\mathbf{x} \cdot \hat{\mathbf{e}}_{\mathbf{\Delta}} - v_0 t) )
\end{equation}
where $m_0 = \sqrt{a/b}$ gives the amplitude of the wave and $\gamma = \sqrt{a/2D}$ gives its width. Eq.~\ref{eq:soliton} defines a 1-dimensional domain wall moving in the direction $\hat{\mathbf{e}}_{\mathbf{\Delta}}$ with velocity $v_0$. This steady-state solution is also the lowest energy deformation for the equilibrium mean-field Ising model \cite{Kardar2007}, and therefore arises frequently from random initial conditions (Supplemental Movie 5).

\begin{figure}
    \centering
    \includegraphics[width=.8\columnwidth]{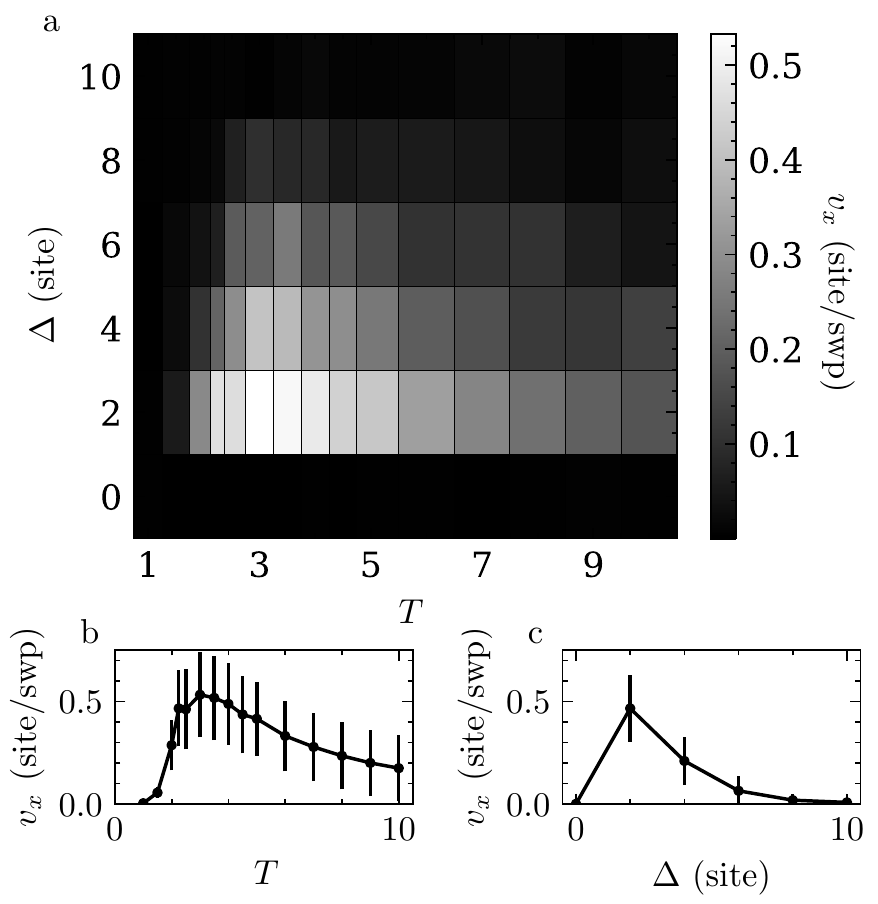}
    \caption{\textbf{Temperature and Offset Control Flow.} (a) Phase diagram of flow speed $v_x$ for changes in offset $\Delta$ and temperature $T$ for lattices of size $L=100$. (b) Flow speed $v_x$ as a function of temperature for $L=100$ with $\Delta = 2$. (c) Flow speed $v_x$ as a function of $\Delta$ for $L=100$ with $T = 2.25$. All error bars represent one standard deviation.}
    \label{fig:phase}
\end{figure}

To determine the extent to which this apparent flow can be controlled, we measure the average flow speed $v_x$ within Monte Carlo simulations as a function of both $\Delta$ and $T$ (Fig.~\ref{fig:phase}a). We observe non-trivial behaviors of the wave speed; wave speed is maximized for $T\sim T_c$ and small non-zero values of $\Delta$. From this maximum, wave speed decreases with increases in $\Delta$ or increases in $T$ 
(Fig.~\ref{fig:phase}b-c). This is in contrast with the monotonic advection speed found in the mean field model (Eq.~\ref{eq:meanfieldspeed}) suggesting the importance of fluctuations in wave propagation. When $T > T_c$, in the regime where our mean-field theory is most accurate, we indeed find that simulations obey the $T^{-1}$ scaling behavior predicted by Eq.~\ref{eq:meanfieldspeed} (Supplemental Fig.~4). Consistent with the 2D Ising model, Monte Carlo simulations with $\Delta = 0$ have no measurable wave speed at any temperature; waves only exist for $\Delta \neq 0$.

\section{Discussion}
A simple alteration of the Ising model, the incorporation of these non-reciprocal interactions which break parity symmetry, leads to traveling spin waves. The directionality of these waves is controlled by the spin-spin interaction, and the speed is a consequence of both the geometry of the interaction and the proximity of the system to its critical temperature. From Monte Carlo simulations, the peak in wave speed near the 2D critical temperature and the lack of wave formation in an equivalent 1D Ising model reinforce that this behavior is intimately linked to the existence of a phase transition (Supplemental Fig. 5). This specific form of non-reciprocity leads to parity-breaking advection in the mean-field equations in all spatial dimensions, as argued on phenomenological grounds and computed from microscopics.

There is ongoing debate about the appropriate usage of Monte Carlo methods for non-equilibrium systems~\cite{levis_clustering_2014, di_pietro_martinez_out--equilibrium_2020, klamser_kinetic_2021}. To explore this further, we repeated our simulations using Glauber dynamics \cite{Glauber1963} which have been used previously to align spins within the active Ising model \cite{solon_revisiting_2013}. We notice that the measurable quantities in Fig.~\ref{fig:fig1} and the subsequent phase transition temperatures are sensitive to whether we utilize the Metropolis or Glauber algorithms for $\Delta \neq 0$ (Supplemental Fig. 6). This is expected as the individual energy, $E_i$, is not a true energy functional when $\Delta \neq 0$, meaning that the steady-state distribution is not simply a function of $E_i$, but also depends on the specific dynamics used. However, we continue to observe a non-monotonic flow speed for non-zero offsets that is qualitatively similar for the two algorithms, suggesting that this apparent flow and propagation of spin fluctuations is a robust feature of the model (Fig.~\ref{fig:phase}b, Supplemental Fig. 6).

We confirm that the existence of these traveling waves is neither a consequence of finite system size nor periodic boundary effects (Supplemental Movies 6,7, and 8). Further, these waves are not a product of intrinsic geometric frustration, as is seen in the triangular Ising model~\cite{wannier_antiferromagnetism_1950}, or proximity to a boundary (Supplemental Movie 8). 

Ultimately, this simple lattice-based model serves as a simplification of the 2D Vicsek model where swimmers are 1) confined onto a grid (i.e. the swimming speed is zero and local density fluctuations are not allowed) and 2) constrained to orient up or down. These simplifications allow us to clearly observe how a type of non-reciprocal interaction propagates fluctuations in the polarization within a system. Importantly, these spin waves are a non-equilibrium effect that is isolated from the non-equilibrium effects of the propulsive energy input ubiquitous to active matter~\cite{vicsek_novel_1995, dadhichi_nonmutual_2020}.

Our use of an offset $\Delta$ and a constant interaction kernel is inspired by so-called `vision-cones'~\cite{durve_first-order_2016, durve_active_2018,chen_fore-aft_2017} which restrict the geometry of active particle interactions; particles can only see a fraction of the entire space, governed by the vision cone angle. Unlike our model, a vision-cone model would have heterogeneous, spin-dependent offset kernels. 
At the mean-field level, we can capture this spin-dependence by considering a variation on our dynamics where the offset is defined locally as $\boldsymbol{\Delta}_i = s_i \Delta_0 \hat e_x$, where $\Delta_0$ is constant for all spins. The mean-field dynamics derived from these microscopic interactions still contain an advection term, as in Eq.~\ref{eq:pitchfork+advection}, but the speed is now $v_0 \propto -m \Delta_0$. This gives a non-linear advection term of the form found in Burger's equation, $m \nabla m$ (Supplemental Note A4). Perturbations around the disordered state are purely diffusive (Supplemental Fig. 7), while perturbations around an ordered state $\pm m_0$ decay while propagating in a direction set by the sign of the ordered state. (Supplemental Fig. 8). This is consistent with a previous study of an XY-model with vision-cone interactions, which found no time-dependent phase but did find transient translations of defects (see Fig. 4 in \cite{Loos2022}).

In order to capture the non-monotonic relationship between the wave speed and temperature, we should not use the mean-field dynamics with temperature dependent coefficients, but rather use noisy Model A dynamics with a $\phi^4$ free energy \cite{Hohenberg1977} supplemented with advection. While appearing superficially similar, the two models have a fundamental difference. The mean-field dynamics Eq.~\ref{eq:pitchfork+advection} contain all the temperature dependence in their coefficients, while Model A has constant coefficients and the temperature dependence is in the strength of the noise. The mean-field dynamics only correctly capture the onset of magnetization, namely $m \sim |T - T_c|^{1/2}$. On the other hand, Model A possesses all the same critical phenomena as the Ising model \cite{Hohenberg1977}, at the cost of losing connection between the microscopic parameters and the phenomenological coefficients. However, including this noise would allow for the wave speed to interact with fluctuations with specified correlation lengths, which we hypothesize to be the origin of the non-monotonic relationship of $v(T)$ observed in Fig.~\ref{fig:phase}.

Finally, we note that this work is consistent with previous works that show a motility change upon the introduction of non-reciprocal interactions~\cite{lavergne_group_2019, costanzo_milling-induction_2019, you_nonreciprocity_2020, saha_scalar_2020}. We hope that our model will help towards decoupling the thermodynamic effects of anisotropic interactions and propulsive energy usage in collective dynamics. 


\section{Acknowledgements}
APT acknowledges support from the M.J. Murdock Charitable Trust Award ID 201913717 and the National Science Foundation Award ID 2137509. DSS acknowledges support from the Kadanoff-Rice Postdoctoral Fellowship. APT thanks John J. Williamson, and Vikrant Yadav for helpful comments. DSS thanks Michel Fruchart for help with the mean-field dynamics and, together with Yael Avni and David Martin, for discussing the distinction between using system energy and individual spin energy in Monte-Carlo simulations.
APT conceived the project. APT, DSS, and AP performed simulations and analysis. DSS performed theoretical analysis. APT, DSS, and AP wrote the manuscript.

\appendix
\section{Monte Carlo Algorithms}\label{app:MC_methods}
The system starts in a configuration given by the vector $\mathbf{s}$ and is contact with a heat bath at temperature $T$. In a single sweep, $N$ lattice sites are randomly selected to be flipped, going to a state $F_i \mathbf{s}$, where $F_i$ is an operator that takes $s_i \mapsto -s_i$
\begin{equation}
    (F_i \mathbf{s})_j = s_j(1 - 2\delta_{ij}).
\end{equation}
For each site $j$, the difference $\Delta E_j = E_j(F_j \mathbf{s}) - E_j (\mathbf{s})$ is calculated, where $E_j(F_j \mathbf{s})$ is the individual energy of spin $j$ after the flipping spin $j$ and $E_j(\mathbf{s})$ is the individual energy of spin $j$ prior to the spin-flip (Eq.~\ref{eq:spinEnergy}). We stress that, generically, the change in the individual energy is \textit{not} the change in the system energy, $E(\mathbf{s}) = \sum_j E_j(\mathbf{s}) / 2$. To illustrate the difference between the two, one can explicitly calculate the change in system energy $\Delta E$ from flipping spin $i$,
\begin{equation}
\begin{aligned}[b]
    \Delta E
    &=
    E(F_i \mathbf{s}) - E(\mathbf{s})
    \\
    &=
    \dfrac{1}{2} \left(
    -\sum_{mn} J_{mn} (F_i \mathbf{s})_m (F_i \mathbf{s})_n
    + \sum_{mn} J_{mn} s_m s_n  \right)
    \\
    &= 
    s_i \sum_{j} \left( J_{ij} + J_{ji} \right)(1 - \delta_{ij}) s_j.
\end{aligned}
\end{equation}
By contrast, we can calculate $\Delta E_i$ due to flipping spin $i$ using Eq.~\ref{eq:spinEnergy},
\begin{equation}
\begin{aligned}[b]
    \Delta E_i
    &=
    E_i(F_i \mathbf{s}) - E_i (\mathbf{s})
    \\
    &=
    -(F_i \mathbf{s})_i \sum_{j} J_{ij} (F_i \mathbf{s})_j + s_i \sum_j J_{ij} s_j
    \\
    &=
    2 s_i \sum_{j} J_{ij}(1 - \delta_{ij}) s_j. \label{eq:spinEnergyChange}
\end{aligned}
\end{equation}
We therefore see that $\Delta E = \Delta E_i$ when $J_{ij} = J_{ji}$, i.e. when the dynamics are reciprocal.

Using the Metropolis-Hastings algorithm, a proposed spin flip is accepted with probability
\begin{equation}
    P_\mathrm{MH}(\mathbf{s} \to F_i \mathbf{s}) = \mathrm{min} \left[ 1, e^{-\beta \Delta E_i} \right].
\end{equation}
Using Glauber dynamics, a proposed spin flip is accepted with probability
\begin{equation}
    P_\mathrm{G}(\mathbf{s} \to F_i \mathbf{s}) = \dfrac{1}{1 + e^{\beta \Delta E_i}},
\end{equation}
where $\Delta E_i$ is given by Eq.~\ref{eq:spinEnergyChange} and $J_{ij}$ is given by Eq.~\ref{eq:bigJ}.

\section{Monte Carlo Measurements}\label{app:MC_Measure}

All Monte Carlo simulations are done on a 2D lattice with periodic boundary conditions. Each spin is initialized to $+1$ and evolved in the presence of a heat bath at a temperature $T$ using the Metropolis Monte Carlo method outlined in Giordano and Nakanishi~\cite{giordano_computational_2006} for 10000 sweeps to reach steady-state. When making measurements in Figs.~\ref{fig:fig1} \&~\ref{fig:Tc_measure}, we build an ensemble of measurements using the following scheme. Near $T_c$ (i.e. $2.1<T<2.5$), we take measurements of the system evolving in time. At steady-state, 5000 additional sweeps are used to calculate the correlation function $c(t)$ of the system magnetization $m$ which evolves in time $t$ ~\cite{newman_monte_1999}
\begin{equation}\label{eq:correlation}
            c(t) = \frac{1}{t-t_{max}}\sum_{t'=0}^{t_{max} - t}m(t')m(t'+t) - \frac{1}{t-t_{max}}\sum_{t'=0}^{t_{max} - t}m(t')\times \frac{1}{t-t_{max}}\sum_{t'=0}^{t_{max} - t}m(t'+t).
\end{equation}
We find that the decay of $c(t)$ can be described by a correlation time $\tau$, which in general depends on lattice size and temperature. We take this steady-state configuration generated after 15000 sweeps, and we continue to evolve it for additional sweeps only taking measurements of the magnetization every $3\tau$ sweeps until we have accumulated 1000 independent measurements. This method is also done far from $T_c$ (i.e. $T<2.1$ and $T>2.5$) in Fig.~\ref{fig:Tc_measure}a. Far from $T_c$ in Fig.~\ref{fig:fig1} and the remainder of Fig.~\ref{fig:Tc_measure}, we build an ensemble of 100 independent measurements instead by taking 100 independent lattices and evolving them for 10000 sweeps in the presence of a heat bath at temperature $T$.

\section{Cumulant Intersection Method} \label{app:binder}
The Binder cumulant $U_L(T)$ is calculated for three lattice sizes $L$ and is fit to a second-order polynomial. The intersection point is determined by the temperature that minimizes the total distance between these three lattice sizes. Lattice sizes are chosen such that a unique cumulant crossing point is observed. For offsets $\Delta=0,2,4,8,10$, we use lattices with $L=[16,32,64],[16,32,64],[32,50,64],[50,64,100],\mathrm{\ and \ }[64,80,100]$, respectively.

\section{Image Analysis}\label{app:image_analysis}

For image analysis techniques, each lattice is rendered as an image where an up/down spin corresponds to a white/black region of pixels, respectively. Each $100\times100$ lattice is converted to a $380 \times 380$ pixel image. Kymographs (spatio-temporal plots) of these images are created using FIJI ~\cite{schindelin_fiji_2012}. Each line of the kymograph comes from an average pixel projection of a 17 pixel tall region spanning the width of the image. These lines are stacked atop corresponding frames in the
image sequence to generate the full kymograph. 

Rendered images are processed with a Gaussian filter with a radius of 4 pixels, and 500 frames are analyzed using particle image velocimetry (PIV). PIV segments two consecutive frames of the rendered video into grids. A velocity vector is drawn to quantify the center of mass displacement within each grid, and we quote all velocity vectors for all grids in all frames. PIV is performed with the OpenPIV Python library using a window size of 64 pixels and an overlap area of 32 pixels ~\cite{liberzon_openpivopenpiv-python_2021}.

\section{Mean-field Dynamics}\label{app:dedalus}
The Python package Dedalus \cite{Burns2020} was used to simulate the mean-field dynamics, Eq.~\ref{eq:pitchfork+advection} using pseudo-spectral methods. We assume periodic boundary conditions on a flat geometry and solve the PDE using 256 Fourier modes over a domain of size $L = 100$.

\section{Bibliography}
\bibliography{lib_July2022.bib}

\end{document}